\documentclass[aps,prb,twocolumn,groupedaddress]{revtex4-1}

\usepackage{graphicx}
\usepackage{color}
\begin{document}


\title{Magneto-optical Kerr Effect Studies of Square Artificial Spin Ice}


\author{K.~K.~Kohli}
\author{Andrew~L.~Balk}
\author{Jie~Li}
\author{Sheng~Zhang}
\author{Ian~Gilbert}
\author{Paul~E.~Lammert}
\author{Vincent~H.~Crespi} 
\author{Peter~Schiffer}
\author{Nitin~Samarth}
\email[]{nsamarth@psu.edu, kkohli@psu.edu}
\affiliation{Department of Physics and Materials Research Institute, The Pennsylvania State University, University Park, Pennsylvania 16802, USA}


\date{\today}

\begin{abstract}
We report a magneto-optical Kerr effect (MOKE) study of the collective magnetic response of artificial square spin ice, a lithographically-defined array of single-domain ferromagnetic islands. We find that the anisotropic inter-island interactions lead to a non-monotonic angular dependence of the array coercive field. Comparisons with micromagnetic simulations indicate that the two perpendicular sublattices exhibit distinct responses to island edge roughness, which clearly influence the magnetization reversal process. Furthermore, such comparisons demonstrate that disorder associated with roughness in the island edges plays a hitherto unrecognized but essential role in the collective behavior of these systems.
\end{abstract}


\maketitle


Arrays of lithographically fabricated single-domain nanoscale ferromagnets can be designed to frustrate inter-island magnetostatic interactions, in analogy to the spin-spin interactions in frustrated magnetic materials~\cite{nature1}. A wide range of interesting behavior~\cite{int1, int2} can be observed in these artificial frustrated magnets by tuning the geometry of square~\cite{square, square2, sqref, square3}, triangular~\cite{triangular, tri2}, hexagonal (and kagome)~\cite{brick, kagome, honeycomb, kagome2, kagome3, kagome4, kagome5, kagome6, honeycomb2, njp, kagref} and brickwork lattices, as well as isolated clusters~\cite{clusters}. In particular, the local moment behavior of these systems has shown that they are good realizations of ice models, and a range of studies have examined monopole excitations and drawn upon close analogies with the pyrochlore spin ice materials.  

The artificial spin ice systems, however, are subject to the limitations of lithography that introduce disorder in the form of variation of features at the nanometer scale. While this disorder is quite different from the point or line defects intrinsic to an atomic lattice, it has the potential to influence the physics of these systems in interesting ways. The manifestations of disorder have been extensively studied in the pyrochlores and other frustrated magnetic materials, and off-stoichiometry and other structural disorder has been demonstrated to lead to changes in the the low temperature collective spin states, zero-point entropy, and other exotic phenomena associated with frustration ~\cite{disorder1, disorder2, disorder4}. Despite the demonstrated importance of disorder in frustrated magnetic materials, there has not yet been a detailed examination of the effects of disorder on artificial spin ice or related systems.

We report magneto-optical Kerr effect (MOKE) studies of square artificial spin ice, with {\it in situ} measurements of the global lattice magnetization. These measurements, which complement the numerous studies of local moments in artificial spin ice, reveal a coupling between the collective behavior of this system and disorder in the shape of the islands. We also demonstrate a method through which the disorder can be considered within micromagnetic simulations, and we provide a simple model through which the interwoven effects of disorder and magnetostatic interactions can be understood. The consideration of disorder opens possibilities for closer comparisons between the artificial spin ice systems and theoretical models, as well as for more detailed comparisons with the pyrochlores, for which only the behavior averaged over an extensive lattice can be examined.  

\begin{figure}
\includegraphics{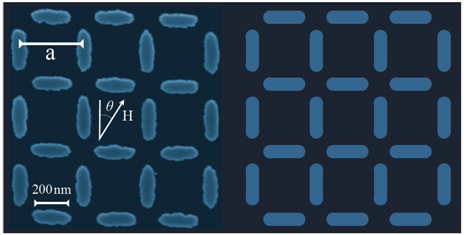}
\caption{\label{FigSEM} (color online) An SEM image of a square array at $a=320$~nm lattice spacing with the magnetic field at an angle $\theta$ to an array axis, compared to an array of 24 ideal stadium-shaped islands, also at a 320~nm spacing.}
\end{figure} 

Our square-ice arrays were fabricated using electron beam lithography, as described elsewhere~\cite{nature1}. The samples consist of 220~nm long, 80~nm wide and 25~nm thick permalloy ($81\%$Ni, $19\%$Fe) islands with lattice constants ranging from 320~nm to 880~nm for the square arrays. The array geometry that was programmed into the electron beam writer is shown in Figure~\ref{FigSEM} on the right. Scanning electron microscopy (SEM) of the resulting nanomagnet array (shown in Figure~\ref{FigSEM} on the left) revealed a surface roughness of $\pm 4.3$~nm on the edges of the islands, defined as the standard deviation of SEM image edges from an ideal island edge. As demonstrated previously through magnetic force microscopy (MFM) studies~\cite{nature1}, the islands are sufficiently small and elongated to generally behave as single-domain ferromagnets, with the strong shape anisotropy directing the island moments along their long axes. A magnetic field applied {\it in situ} within a MOKE magnetometer (in the longitudinal geometry~\cite{Kerr}) enables acquisition of full hysteresis loops during magnetization reversal. The $50~\mu$m spot of an s-polarized HeNe laser was focused onto the arrays via an optical microscope, with the sample mounted on precision \textit{XY} translation stages modified to allow sample rotation in the magnetic field. The reflected beam was polarization analyzed using lock-in detection to extract the sample-generated magnetization-dependent Kerr rotation. For an array with a lattice spacing of 320~nm about $40,000$ islands were simultaneously probed, while only $\sim$5000 islands were probed for the largest lattice spacing, 880~nm. The field was incrementally ramped, in steps of 10~Oe near the switching fields and 100~Oe otherwise. The raw data were smoothed by local second-order polynomial regression around each point using a Savitzky-Golay filter~\cite{Savitzky}.
 
\begin{figure}
\includegraphics{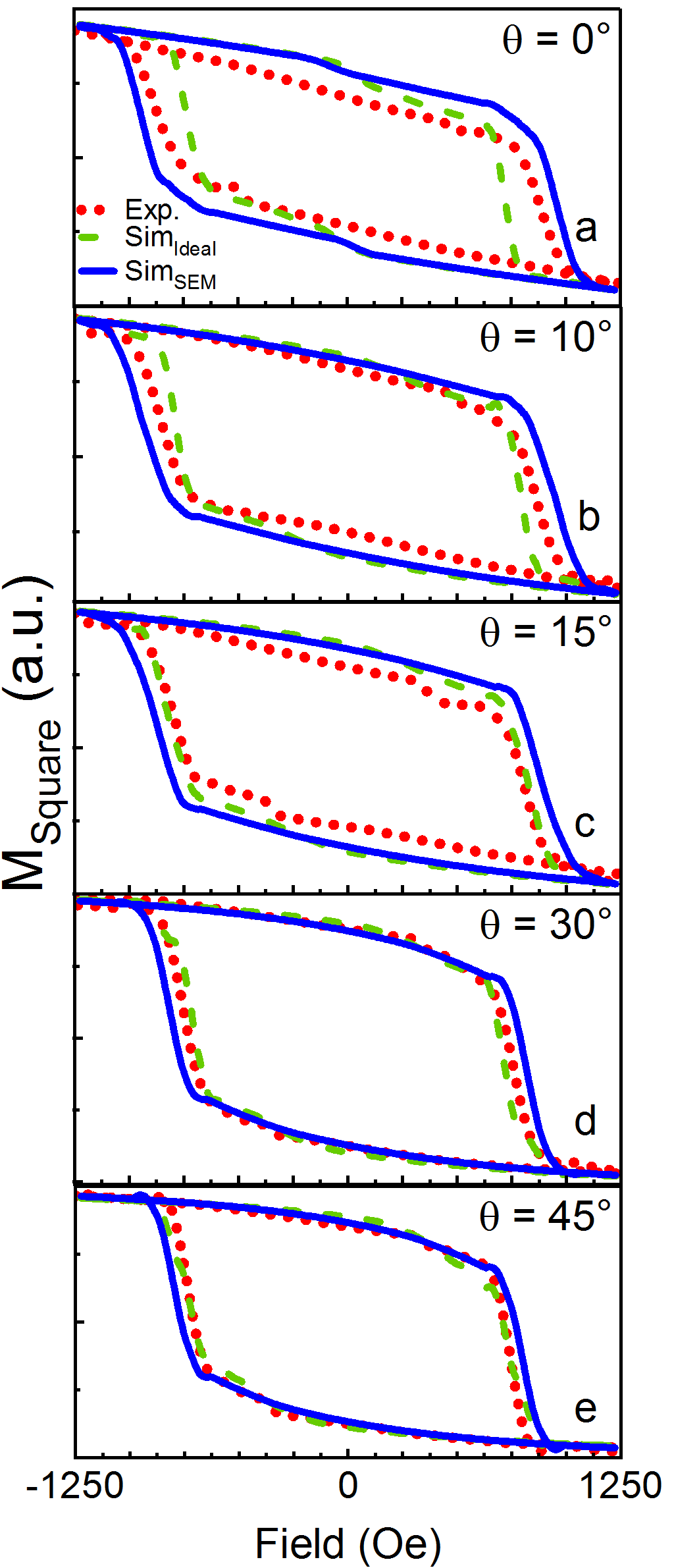}
\caption{\label{Fig2} (color online)(a)-(e) Hysteresis loops comparing MOKE measurements (red dotted line) to simulations at $\theta = 0, 10, 15, 30, 45^{\circ}$ for square arrays with 320~nm lattice spacing. The green dashed line and blue solid line show simulations for ideal and experimental island shapes, respectively. Note the significant changes with angle that are summarized in Figure~\ref{Fig3}.}
\end{figure} 

\begin{figure}
\includegraphics{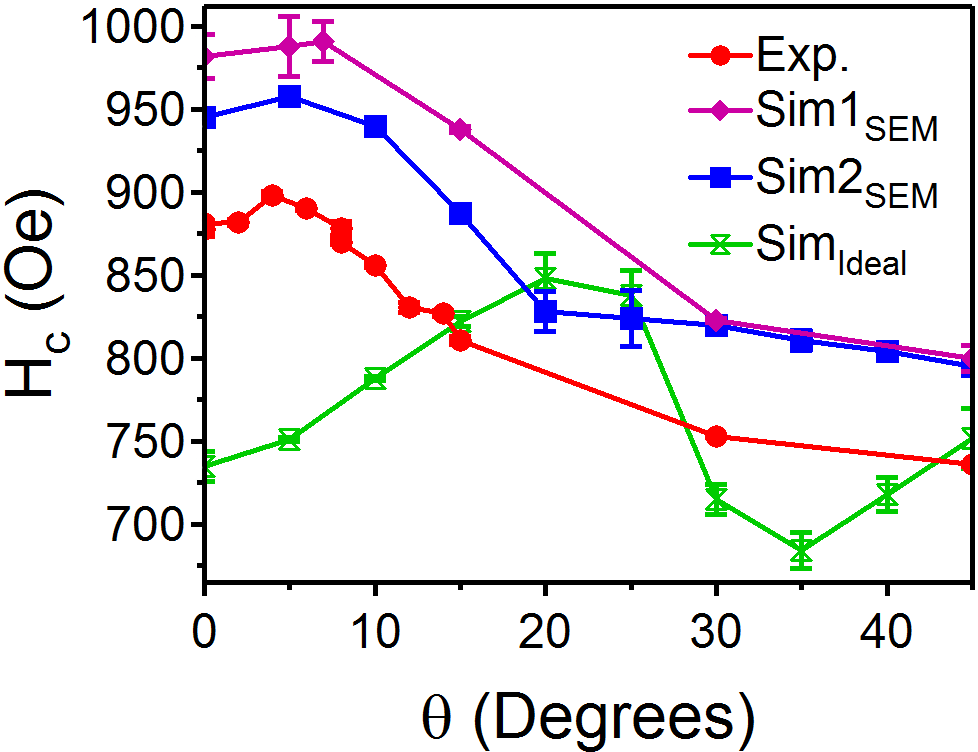}
\caption{\label{Fig3} (color online) Coercivity as a function of angle for a square array with 320~nm lattice spacing. Simulations were run on a 24-island cluster similar to that of Fig.~\protect\ref{FigSEM} for both ideal and SEM-derived island shapes. Sim1 and Sim2 refer to two different simulations run for SEM images of different islands.}
\end{figure}

Figure~\ref{Fig2} shows hysteresis loops obtained from arrays with 320~nm lattice spacing for different angles $\theta$ of the magnetic field with respect to one of the primary axes of the square lattice (as defined in Fig.~\ref{FigSEM}). The hysteresis loops have a strong dependence on $\theta$ in both shape and coercive field $H_c$. For certain angles, the steepest part of the hysteresis curves occurs away from $M=0$, such that $M(H)$ near $M=0$ is only weakly sloped. Hence a definition of $H_c$ in terms of the magnetization zero-crossing is problematic in this regime, and we instead define the coercive field as the field of maximum slope in $M(H)$, a definition which enables a more consistent and unbiased estimate for our samples. We used a range of smoothing parameters in the Savitzky-Golay filter to estimate effective error bars for simulations; the error in the experimental data was determined by the uncertainty in the peak fit. The extracted values of $H_c$($\theta$) shown in Figure~\ref{Fig3} reveal a strong and {\it non-monotonic} angular anisotropy, with a local maximum near $\theta = 5^{\circ}$ and a minimum at $\theta = 45^{\circ}$. For all lattice spacings, the qualitative features of the coercivity are symmetric around $\theta = 45^{\circ}$, as expected due to the square lattice symmetry. We find a slight asymmetry in magnitude that we attribute to the rastering intrinsic to e-beam lithography, but the key features for $\theta$ and ($90-\theta$) remain qualitatively equivalent.

We simulated the arrays' magnetic response using the NIST OOMMF (object oriented micromagnetics framework) code~\cite{OOMMF} for an array of 24 islands in the geometry shown in Fig.~\ref{FigSEM}. We used an OOMMF cell size of $5\times5\times5$~nm$^3$, comparable to the exchange length of permalloy~\cite{ex1, ex2}. The saturation magnetization ($860\times10^3$~Am$^{-1}$) and exchange constant ($13\times10^{-12}$~Jm$^{-1}$) are standard literature values~\cite{ferro}. Fig.~\ref{Fig2} plots the resulting $(M$-$H)$, smoothed using the same methods as for the experimental data, and Fig. ~\ref{Fig3} displays the resulting $H_c(\theta)$. 

The simulation does {\it not} match the experimental data when using ideal stadium-shaped islands, especially the small minimum near $\theta = 0$. This mismatch is consistent with previous studies showing that nominally identical magnetic nanostructures can have significant variations in their switching fields, presumably due to shape variations during fabrication~\cite{cow1}.  Prior micromagnetic simulations have incorporated such effects by using randomly generated edge profiles, periodic removal of edge elements, or edge roughness models based on experimental observation~\cite{cow3}. We incorporated island edge roughness in the simulation by basing the island shapes directly on SEM images like that of Figure~\ref{FigSEM} (the pixel size of the SEM images was always considerably smaller than the OOMMF cell size - typically around $2\times2$~nm$^2$). Micromagnetics simulations of isolated islands show that this degree of shape disorder is sufficient to vary the coercivity of an isolated island by $\sim \pm 80$ Oe. As shown in Figs.~\ref{Fig2} and \ref{Fig3}, the simulated magnetic response of the island arrays with SEM-inferred shape disorder successfully reproduces the overall shape of the experimental curves, including the local maximum in $H_c$ near $\theta = 5^{\circ}$  and monotonic fall-off at higher angles.  

Simulations with other images or with different arrays of simulated islands gave qualitatively consistent results. We varied both the array boundaries and size of the arrays to test the sensitivity, and we found only a $1\%$ change in coercivity when increasing the total number of islands in the simulation from $24$ to $40$ and less than $3\%$ between $12$ islands and $40$ islands. We also observed less than a $1\%$ change in coercivity at $\theta = 0^{\circ}$ when we simulated a cluster of 24 islands chosen with different boundaries from those shown in Fig.~\ref{FigSEM}. The main residual discrepancy between simulations and the experimental data is a roughly uniform vertical shift in the simulation to higher overall fields.

\begin{figure}
\includegraphics{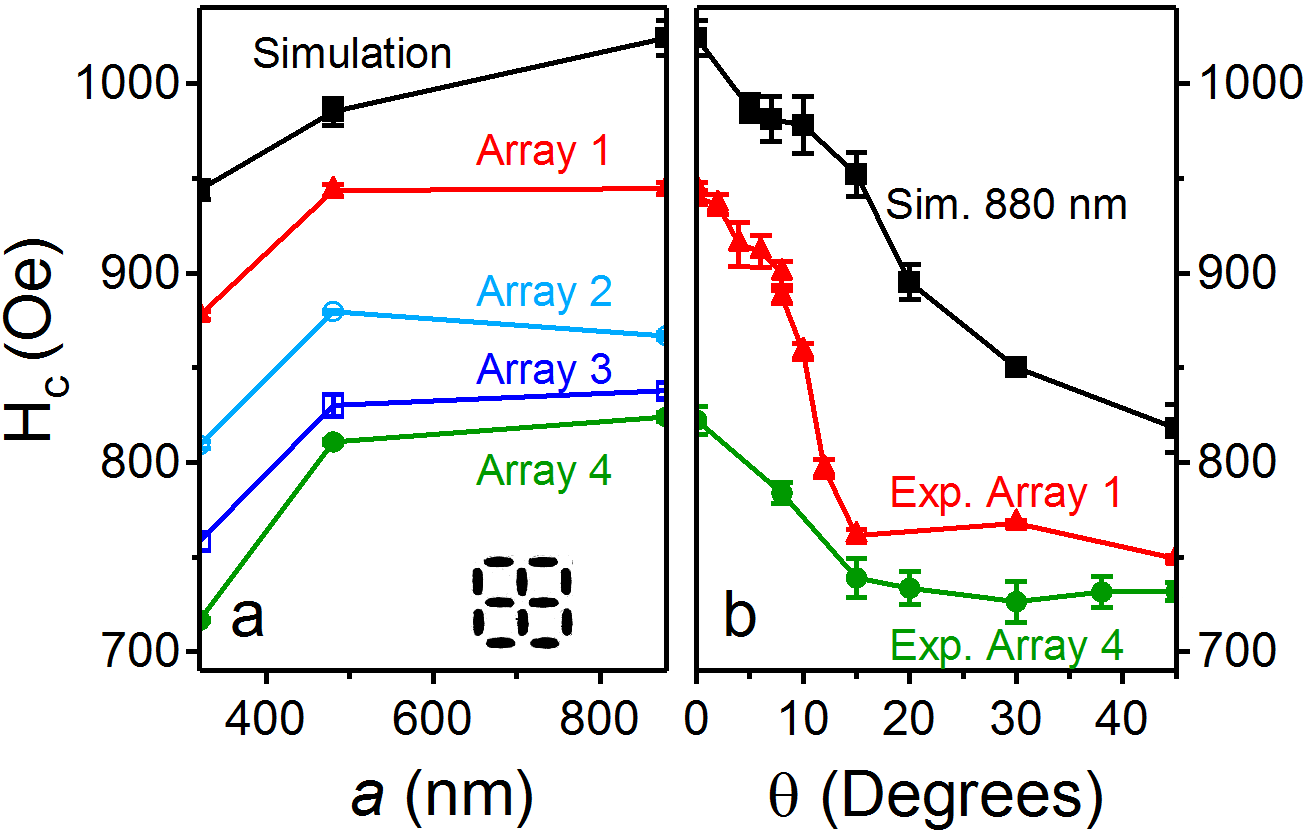}
\caption{\label{Fig4} (color online) (a) Variation of coercivity at $\theta = 0$ as a function of lattice spacing for square arrays. Measurements on various samples (arrays 1-4) are compared with micromagnetic simulations (black solid squares), comparing simulations of the same island outlines but with different spacing. Error bars for experiment are smaller than the data points. (b) Experiment and simulation for square arrays as a function of angle for the largest lattice spacing of 880~nm. The local maximum in $H_c$ near $\theta = 5^{\circ}$ that appeared for 320~nm is missing for the larger lattice spacing. Simulations were run on a 12-island array as indicated in the inset, with an appropriately dilated lattice spacing between the SEM islands. }
\end{figure}

To elucidate the role of island-island magnetostatic interactions in the collective properties of the arrays, we also measured hysteresis loops for arrays with larger lattice spacings of $a = 480, 880$~nm, for which the interactions are much weaker than in the 320~nm array discussed above ~\cite{nature1}. As shown in Fig.~\ref{Fig4}(a), the coercivity as a function of lattice spacing at $\theta = 0$ increases with increasing lattice spacing for each of three different sets of arrays (two from one processing run and a third from another). Similar behavior from Ref.~\cite{demag} (shown as array 3) and from simulations also confirms this effect. These results clearly demonstrate that interactions do change the coercive field. As shown in Fig.~\ref{Fig4}(b), the angle-dependence of  the coercivity at $a = 880$~nm shows no local maximum in $H_c$ near $\theta = 5^{\circ}$, neither in experiment nor in simulations (using the twelve-island array shown in the inset with SEM-derived island shapes)~\cite{480also}. Hence, we can conclude that the feature at small $\theta$ for the 320~nm lattice is associated with both island shape disorder and inter-island interactions and demonstrates the interplay between interactions and disorder effects in these systems. 

We now consider how the interplay between disorder and the inter-island interactions lead to the increase in the coercive field with island separation at $\theta = 0$ and the local maximum in $H_c$ near $\theta = 5^{\circ}$ for the 320~nm lattice constant. For small values of $\theta$, we can qualitatively understand the effect of interactions among the sublattice of ``vertical'' islands, aligned with the field, by careful examination of the micromagnetic simulations. When the field is swept through $H_c$, the magnetization of an island reverses suddenly when the total field (the sum of the external field and the field from other islands) reaches a critical value specific to that particular island. These moment reversals of the vertical islands account for most of the change in the net magnetization of the system along the steep parts of the hysteresis curves. The effect of an island that has reversed to align with the external field is to enhance the external field near that island, while an island that has not reversed acts to reduce the magnitude of the total field acting on its neighbors. The enhancement is stronger than the reduction before moment reversal; just before reversal, an island has complex magnetization profile at the tips, whereas after reversal, its magnetization field is ``stretched out'' with strong poles. This asymmetry allows islands with lower intrinsic coercivity to initiate cascades of reversals \cite{Kohli_sup}, thus decreasing the coercivity of the entire array. The cascade phenomenon, which is naturally affected by the island edge roughness, is clearly seen in the simulations and it explains at least some of the decrease in coercive field at small lattice spacing for all field angles. Indeed, simulations show that approximately $60\%$ of the total decrease in the coercivity at $\theta = 0$ due to interactions originates entirely from interactions among the vertical islands. 

The sublattice of ``horizontal'' islands, nearly perpendicular to the applied field, plays a subtle role in altering the coercivity. OOMMF simulations of a 320~nm array of only ``vertical'' islands shows some flattening of the angle-dependence of the coercivity, relative to the coercivity for the 880~nm array at small angle, but the maximum is still at $\theta = 0$. We therefore conclude that the shift of the maximum $H_c$ to a small non-zero angle, $\theta_{\max}$, is associated with the effects of the horizontal islands. A possible explanation is that $\theta_{\max}$ is the angle at which the net effective horizontal field on the vertical islands is zero. According to OOMMF simulations \cite{Kohli_sup}, at field angles of order $\theta = 5^{\circ}$ or greater, the magnetizations of horizontal islands rotate in unison as the field is swept; near $H_c$, the horizontal islands are all magnetized nearly along their easy axes with the same orientation. This creates an effective horizontal field acting on the vertical islands that opposes and cancels the horizontal component of the applied field under the right conditions. 

When the external field is at zero angle, OOMMF simulations show \cite{Kohli_sup} that some horizontal islands develop a magnetization to the left, others to the right, yet others go into a vortex state, a process that will be affected significantly by the edge profiles of the islands. In this situation, each vertical island experiences a different effective horizontal field component. On average it is zero, but as each vertical island responds to the local field, the effect on $H_c$ is as though there were a small horizontal component to the field. Thus, a slight dip in the coercivity is expected at $\theta = 0$. If all the horizontal islands could be magnetized in the same direction while the applied field were at $\theta=0$, then the coercivity would be expected to drop yet more. As an experimental test, we temporarily applied a horizontal field to align all the horizontal islands before measurement. Subsequent MOKE measurements showed a decrease of $H_c$ by approximately $10$~Oe, substantially supporting the above picture.

In summary, our data reveal the {\it in situ} collective magnetization as a function of field in the `large-array' limit and the influential role of island shape disorder. The results complement previous local probe studies that have been very powerful in revealing short-wavelength phenomena and have demonstrated that artificial spin ice is a good realization of ice model physics. Given the high energy scales of both the island magnetic anisotropy and the inter-island interactions, our measurements correspond to metastable frozen states similar to those observed at low temperatures in the pyrochlore spin ice materials. As a result, phenomena such as the magnetization steps that are observed in those materials at low temperatures might be expected, although a small {\it in situ} thermalizing fluctuation may be necessary to observe these effects. An interesting extension along these lines would be to combine MOKE techniques with dynamic probes such as microwave excitation of moment reorientation or local thermal excitation to above the ferromagnetic Curie temperature. Such a combination would add a new quasi-thermal aspect to the artificial frustrated spin ice systems and would allow a more direct comparison the pyrochlores.

\begin{acknowledgments}
This research has been supported by the U.S. Department of Energy, Office of Basic Energy Sciences, Materials Sciences and Engineering Division under Award \# DE-SC0005313 and lithography has been performed with the support of the National Nanotechnology Infrastructure Network. We are grateful to Chris Leighton and Mike Ericson at the University of Minnesota for permalloy deposition and helpful discussions.
\end{acknowledgments}

\providecommand{\noopsort}[1]{}\providecommand{\singleletter}[1]{#1}%

\end{document}